# Two-Dimensional Graphene-like BeO Sheet: A Promising Deep-Ultraviolet Nonlinear Optical Materials System with Strong and Highly Tunable Second Harmonic Generation

*Linlin Liu, Congwei Xie\*, Abudukadi Tudi, Keith Butler and Zhihua Yang\**

Linlin Liu, Congwei Xie, Abudukadi Tudi and Zhihua Yang

Research Center for Crystal Materials; State Key Laboratory of Functional Materials and Devices for Special Environmental Conditions; Xinjiang Key Laboratory of Functional Crystal Materials; Xinjiang Technical Institute of Physics & Chemistry, CAS, 40-1 South Beijing Road, Urumqi 830011, China

*Corresponding authors, E-mail: cwxie@ms.xjb.ac.cn (Congwei Xie); zhyang@ms.xjb.ac.cn (Zhihua Yang)

Keith Butler

Department of Chemistry, University College London, Gordon Street, London, WC1H0AJ, UK

**Abstract:** Two-dimensional (2D) materials with large band gap, strong and tunable second-harmonic generation (SHG) coefficients play an important role in the miniaturization of deep-ultraviolet (DUV) nonlinear optical (NLO) devices. Despite the existence of numerous experimentally synthesized 2D materials, none of them have been reported to meet DUV NLO requirements. Herein, to the first time, an experimentally available graphene-like BeO monolayer only formed by NLO-active [BeO$_3$] unit is suggested as a promising 2D DUV NLO material due to its ultrawide band gap (6.86 eV) and a strong SHG effect ($\chi^{(2)}_{22}$(2D) = 6.81 Å·pm/V) based on the first-principles calculations. By applying stacking, strain and twist engineering methods, a number of 2D BeO sheets have been predicted and the flexible structural characteristics endow them with tunable NLO properties. Remarkably, the extremely stress-sensitive out-of-plane $\chi^{(2)}_{15}$(2D) and $\chi^{(2)}_{33}$(2D) (exceptional 30% change) and the robust in-plane $\chi^{(2)}_{22}$(2D) against large strains can be achieved together in AC-, AAC-, AAE, and ACE-stacking BeO sheets under in-plane biaxial strain, exhibiting emergent phenomena uniquely not yet seen in other known 2D NLO materials. Our present results reveal that 2D BeO systems should be a new option for 2D DUV NLO materials.

# 1. Introduction

Deep-ultraviolet (DUV) nonlinear optical (NLO) materials that can produce coherent light below 200 nm by the direct second-harmonic generation (SHG) output from solid-state lasers are of worldwide interest for their advanced applications, such as lithography, precise microfabrication, ultraviolet communication, and high-resolution photoelectric spectroscopy. However, excellent DUV NLO materials are rare in nature due to the difficulty in not only satisfying strict requirements on optical properties such as wide transparency window (i.e., ultrawide band gap, $E_g \geq 6.2$ eV), large SHG, suitable birefringence to make the shortest phase-matching wavelength below 200 nm and so forth[1], but also having the advantages of integration and compatibility due to the trend of integrated miniaturization of multifunctional devices[2]. Dramatically reduced dielectric screening and significantly enhanced Coulomb interactions make 2D NLO materials possess unique electronic, mechanical properties and superior optical properties[3] such as large optical nonlinearities, large light-matter interaction, ultrafast broadband and tunable optical response, and strong excitonic effects, which satisfy the growing needs for miniaturized, integrated, highly efficient and broadband photonic and optoelectronic devices. Furthermore, 2D NLO materials are free of phase-matching bottleneck by virtue of the atomic layer thickness and have unique and tunable properties, leading to potential applications in modern on-chip nanophotonics.

Many 2D materials with large SHG coefficients have been theoretically predicted[4] and experimentally observed[5], including graphene[6], hexagonal boron nitride[7], phosphorene, monochalcogenides (e.g., GaS[8], GaSe[9], GeSe[10], SnS[10], SnSe[10]), metal dichalcogenides (e.g., $TiS_2$[11], $NbSe_2$[12], $MoS_2$[7], $MoSe_2$[13], $MoTe_2$[14], Janus MoSSe[15], $WS_2$[13], $WSe_2$[13], Janus WSSe[15], $PdSe_2$[16], $ZnS_2$[17], $CdO_2$[18], Janus $Bi_2TeSe_2$[19]), metal phosphorous trichalcogenides (e.g., $SnP_2S_6$[20], $SnP_2Se_6$[21], $CuCrP_2S_6$[22], $CuInP_2S_6$[23]), metal oxide dihalides (e.g., $NbOX_2$ [2a, 24], NbOXY[25] (X, Y = Cl, Br, I and X ≠ Y)), 2D $CrX_3$[2c], 2D perovskite[26], α-Sb and α-Bi[27] and so on, which cover transparent regions from infrared to ultraviolet. Moreover, the SHG coefficients in 2D NLO materials can be significantly enhanced by electric control[28], chemical doping[2a], strain[29], interlayer sliding[30], pressure[23, 24c, 24d] and resonant excitonic excitation[31]. However, the vast majority of existing successfully synthesized 2D materials[2a, 2b, 16, 32] do not have sufficiently large DUV ($\lambda \leq 200$ nm) band gap ($E_g \geq 6.2$ eV) or intrinsic noncentrosymmetric structures, and their optical features are not sufficiently diverse to satisfy broadband optoelectronic applications including UV photodetector and transparent transistor.

Despite significant efforts to expand the family of 2D ultrawide band gap semiconductors and explore their ultraviolet region applications in recent years, this field is still in its early stages of development[4b, 4c, 32a, 33]. It is noteworthy that the BeO monolayer (ML) with $sp^2$-hybridized hexagonal lattice[34] only composed of [BeO$_3$] groups has been synthesized experimentally[35] and has excellent physical properties[36]: high air stability[37], hardness, melting point and thermal conductivity (266 Wm$^{-1}$K$^{-1}$ [38], 278 Wm$^{-1}$ K$^{-1}$ [39], and 385 Wm$^{-1}$K$^{-1}$ [40] with different theoretical methods), ultrawide band gap (6.8 eV)[40] and extraordinary elastic modulus (408 GPa) as well as tensile strength (53.3 GPa)[40]. This can be attributed to a significant covalent component in the primarily ionic BeO bond, as revealed by Compton scattering measurements[41]. The coplanar [BeO$_3$] group has been proven to be superior NLO-active functional unit[42], which has the comparable microscopic SHG coefficients with that of the conventional NLO-active [BO$_3$] units[43]. However, to the best of our knowledge, the NLO properties of BeO sheet remain unexplored, despite so much attention paid to 2D BeO systems[44]. The comprehensive estimation and understanding of SHG are essential for expanding the NLO applications to DUV region in nanoscale.

In this work, we systematically investigated the NLO properties of 2D BeO systems by first-principles methods. The 2D BeO ML can exhibit both an ultrawide DUV $E_g$ (6.86 eV) and a strong sheet SHG response ($\chi_{22}^{(2)}$(2D) = 6.81 Å·pm/V). The dependence of the SHG and bandgap on the layer, strain, and twisted angle in 2D BeO sheet are reported. The typical AA-stacking configuration was confirmed to be optimal for enhancing SHG in 2D BeO sheets. Moreover, the band gap of BeO decreases gradually and converges to 6.48 eV (slightly equal to the band gap of the AA-stacked BeO bulk phase of 6.47 eV) with the increasing number of AA stacked layers. Using the effective thickness defined by the electrostatic potential method, the SHG of 2D BeO sheet gradually increases with the number of stacked layers until it reaches the same SHG coefficients as the AA-stacked bulk phase (2.19 pm/V). More interestingly, within the same stress ($\varepsilon$) range -5% ≤ $\varepsilon$ ≤ 5%, the out of plane $\chi_{15}^{(2)}$(2D) and $\chi_{33}^{(2)}$(2D) values in AC-, AAC-, AAE, and ACE-stacking BeO sheets can be linearly changed by ~ 30% because of out-of-plane polarization, while the in-plane $\chi_{22}^{(2)}$(2D) values can even maintain constant. Besides, the SHG coefficients of BeO bilayers decreases with increasing twisted angle, showing high tunability of NLO effects. And the suppressed SHG in twisted BeO bilayers is predominantly attributed to the strong interlayer coupling between the two adjacent MLs. Thus, the discovery of 2D BeO materials constructed by only excellent functional motifs [BeO$_3$] provide an unprecedented opportunity to explore NLO materials and applications in nanoscale.

## 2. Results and Discussion

### 2.1. Structure properties of 2D BeO monolayer.

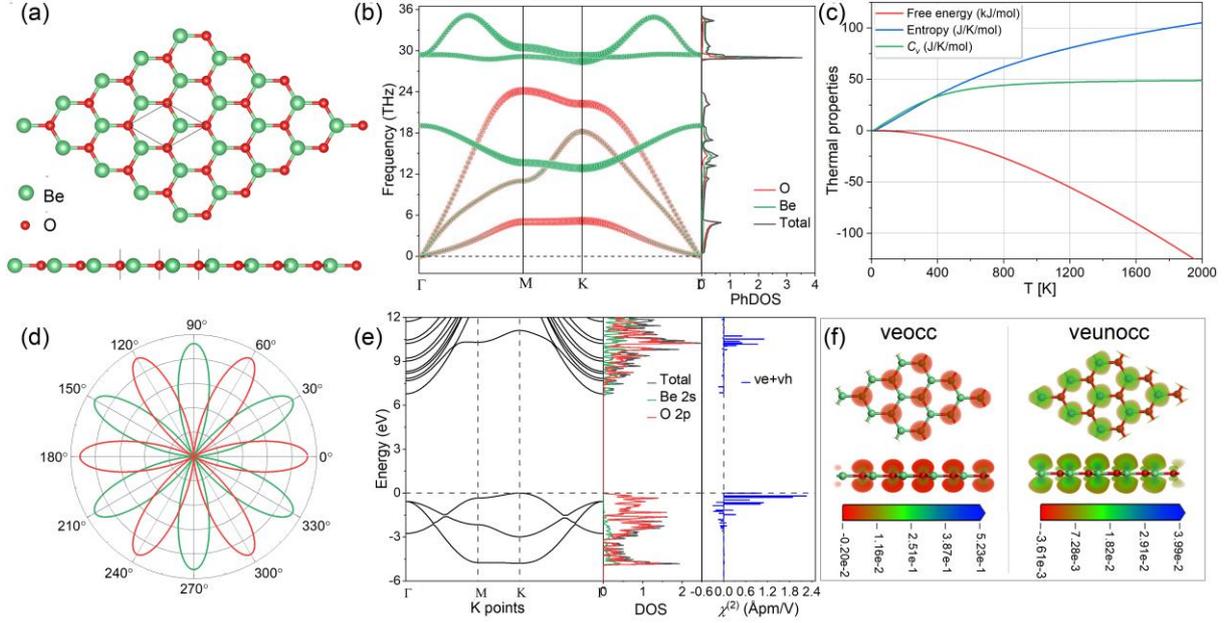

**Figure 1.** (a) Top and side views of optimized structure of BeO monolayer. Green and red spheres represent Be and O atoms, respectively. The hexagonal marked by black line denotes a unit cell. (b) Phonon dispersion and phonon density of states (phDOS) of BeO monolayer. Red and green color represent the contribution of O and Be atoms to the phonon bands, respectively. (c) Thermodynamic properties: Helmholtz free energy (F), entropy (S) and heat capacity ($C_V$) for BeO monolayer. Angle-dependent SHG responses of BeO monolayer (d). The green and red lines represent parallel and perpendicular signals. (e) Band structure, electronic density of states (DOS), and band-resolved SHG contribution of the BeO monolayer, with the corresponding SHG-weighted density (f). ve, vh, veocc and veunocc represent virtual-electron, virtual-hole, virtual-electron-occupied and virtual-electron-unoccupied states, respectively.

The optimized 2D BeO ML possesses non-centrosymmetric space group $P\bar{6}m2$ (no. 187) with lattice constants a = 2.68 Å, which agrees well with the experimental value of 2.65 Å[35a]. The BeO ML is a one-atomic-layered 2D crystal comprised of alternating Be and O atoms by strong $sp^2$ covalent bonds in a flat hexagonal honeycomb lattice like graphene and BN MLs (Figure 1. (a)). The absence of imaginary frequencies throughout the entire Brillouin zone confirms that 2D BeO is dynamically stable. The two highest optical modes are separated from others by a large phonon gap around 25 THz because of mass differences between Be and O atoms. There are three acoustic branches and three optical branches in 2D BeO ML[44], similar to graphene.

The frequency optical modes of 2D BeO ML are governed by Be atom, and the low frequency acoustic branches are mainly dominated by O atom. Detailed analysis of the atom-resolved phonon density of states (phDOS) also reveals that high-frequency phonon modes mainly result from Be atoms, the low frequency phonon modes primarily come from the heavier O atom, whereas in the mid-frequency, both Be and O atoms have contributions to phDOS, as shown in the right pattern of Figure 1.(b).

A series of thermodynamic properties can be derived based on the calculated phonon spectrum as shown in Figure 1(c). With increasing temperature ($T$), entropy ($S$) increase, while Helmholtz free energy ($F$) decreases. As the $T$ increases from 0 K to 2000 K, the $S$ increases from 0 Jmol$^{-1}$K$^{-1}$ to 100 Jmol$^{-1}$K$^{-1}$, but $F$ decreases from 22 kJmol$^{-1}$ to -100 kJmol$^{-1}$. For the heat capacity ($C_v$), with increasing $T$, increases faster and follows the $T^3$ power function dependence at lower $T$ (< 400K)[45]. And then it increases slowly in the higher $T$ region and goes close to 46.07 Jmol$^{-1}$K$^{-1}$. Hence, the BeO ML with high $C_v$ and thermal conductivity[38-40] could be expected to have more resistance to laser damage.

Given the established structural properties of the BeO ML, we next investigated the SHG response properties of 2D BeO sheets. Because the BeO ML belongs to the $D_{3d}$ point group, only one independent nonzero component $-\chi^{(2)}_{xxy} = -\chi^{(2)}_{yxx} = \chi^{(2)}_{yyy}$ is allowed by the limitation of Kleinman symmetry, which are further reduced to $\chi^{(2)}_{22}$. The static SHG susceptibilities of 2D BeO ML are calculated, which are the zero-frequency limit of the SHG susceptibilities. The results demonstrate that the BeO ML exhibits larger $E_g$ (6.86 eV) and $\chi^{(2)}_{22}$(2D) = 6.81 Å·pm/V than those of KBe$_2$BO$_3$F$_2$(KBBF)-like 2D Be$_2$CO$_3$F$_2$ ($E_g$ = 5.20 eV; $\chi^{(2)}_{11}$ (2D) = 5.50 Å·pm/V), slightly lower than those in B$_2$O$_3$ ML only formed by π-Conjugated [BO$_3$] units (12R-B$_2$O$_3$ ML: $E_g$ = 7.20 eV; $\chi^{(2)}_{22}$ (2D) = 16.23 Å·pm/V and 18R-B$_2$O$_3$ ML: $E_g$ = 8.18 eV; $\chi^{(2)}_{11}$ (2D) = 6.86 Å·pm/V, $\chi^{(2)}_{22}$ (2D) = 10.22 Å·pm/V) (Table S1). Such large static SHG susceptibility indicates a strong frequency-doubling effect in BeO ML, which can be attributed to a strong bond in the plane and the high polarization from the asymmetry sublattices of Be and O atoms[44], and indicates that [BeO$_3$] is an effective group for designing two-dimensional DUV NLO materials. Besides, the angular-dependent SHG responses (Figure 1(d)) reveal a six-fold symmetry and highly polarized SHG response of the BeO ML. The maximum value of SHG in the parallel direction appears at 30°, 90°, 150°, 210°, 270° and 330°, and the maximum value in the vertical direction appears at 0°, 60°, 120°, 180°, 240° and 300°. The vertical and parallel directions differ by 60°. Consequently, the SHG coefficients of the 2D BeO ML exhibits an incident angle-dependent anisotropy. To our best knowledge, the 2D BeO ML is the first binary metal oxide that has been successfully fabricated and can be ideal candidates for DUV NLO crystals.

To identify the SHG origin of BeO ML at the (virtual) electronic-transition level, the band structures and density of states (Figure. 1(e)) were calculated, and the band-resolved SHG contribution and corresponding SHG-weighted density (Figure. 1(f)) were obtained. As shown in Figure 1(e), BeO ML possess a DUV $E_g \sim 6.86$ eV, which is larger than those of h-BN (6 eV)[46] and other 2D KBBF family materials $Be_2CO_3F_2$ (5.20 eV)[47] and $AlBeBO_3F_2$ (5.50 eV)[48] ML, and lower than those reported 2D DUV NLO materials $B_2S_2O_9$ (8.63 eV)[4c] and $BeP_2O_4H_4$ (7.84 eV)[4b]. The calculated density of states indicate that the valence-band maximum and conduction-band minimum are mainly composed of O-2p and Be-2s orbitals, respectively. The band-resolved SHG contribution reveals that the virtual-electron processes exhibit main contribution (occupies 99%) to the SHG effect. The SHG-weighted density analyses of the virtual-electron process illustrate the occupied states are mainly the contribution of O-2p orbitals, and the unoccupied states are mainly the contribution of Be-2s orbitals in BeO ML, as shown in Figure 1(f). This result is consistent with the density of states and band-resolved analysis (Figure 1(e)). The electronegative difference between Be and O atoms results in the charge transfer and the high density around O atoms[44]. It is evident that O atoms in 2D BeO ML play a crucial role in contributing to the SHG effect.

## 2.2. Layer dependence of SHG response of 2D BeO sheet.

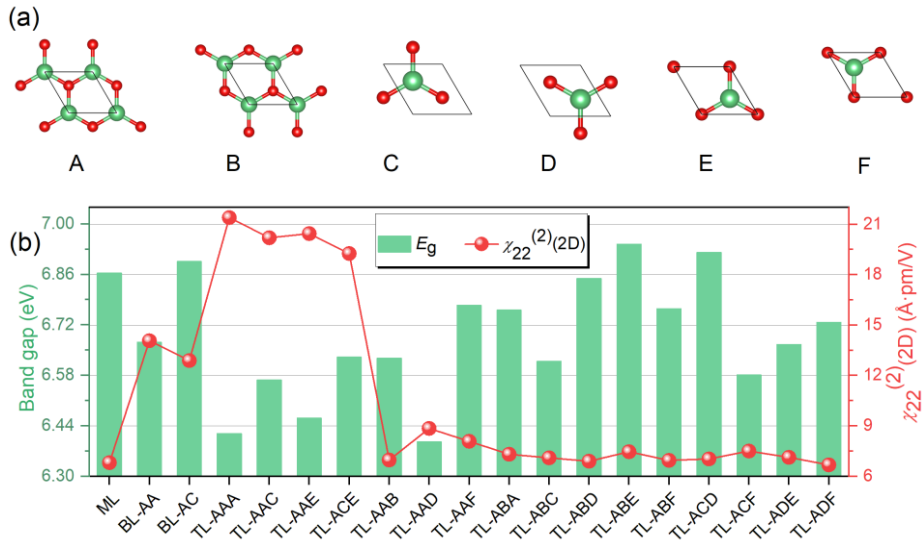

**Figure 2.** (a) Unit cell of 2D BeO with different atomic arrangements. (b) Calculated band gap $E_g$, SHG coefficients $\chi_{22}^{(2)}$(2D) as well as the different layers with different stacking patterns. ML, BL and TL stand for monolayer, bilayer, and tirlayer. BL-AA indicate the BeO bilayer in AA stacking configuration. And the rest is the same.

We further investigate van der Waals stacking induced SHG effects in 2D BeO sheets. All possible bilayer (BL) and trilayer (TL) BeO structures were built based on the various prototypical structures of the ML BeO, considering different stacking patterns and orders (Figure 2 (a)). Among these, the BL BeO sheet shows six stacking configurations and possesses three distinct space groups ($P\bar{6}m2$, $P3m1$ and $P\bar{3}m1$). Their energies range from -7.07 to -7.09 eV/atom, which are much lower than that of the BeO ML (-7.01 eV/atom) (Figure S3). The 2D BeO sheets in BL-AB, -AD, and -AF stacking is centrosymmetric and belong to the space group ($P\bar{3}m1$), displaying no second-order NLO response. The AE-stacking configuration is obtained by flipping AC-stacking by 180° and translating it along the diagonal. Therefore, the subsequent discussion on BL-AB, AD, AE and AF stacking will not be too much. In contrast, the TL BeO sheet has sixteen stacking configurations and belongs to two noncentrosymmetric space group ($P\bar{6}m2$ and $P3m1$). Their energies are in the range of -7.01 to -7.10 eV/atom, which don't show a significant decrease and is comparable to those of BeO BL (Figure S3), indicating an appropriately favorable metastable state.

Although the interfacial van der Waals force is generally quite weak, the stacking patterns, orders and number of layers always play significant roles in determining the ground state (Figure S3), interlayer interaction strength (Figure S4), electronic energy band structure (Figure S5), and static SHG susceptibilities of 2D BeO sheets. From Figure 2(b), the band gap of 2D BeO is not only influenced by the number of stacked layers but also varies with the stacking patterns and orders, especially in the BeO TL. In 2D BeO TL, the bandgap of TL-ABE is up to 6.92 eV, which is the optimal stacking patterns and orders to regulate the bandgap. And the band gaps of TL-ACE (6.85 eV) and TL-ABD (6.85 eV) are comparable to that of 2D BeO ML (6.86 eV), while the bandgap of TL-ADE (6.67 eV) is as large as that of BL-AA (6.67 eV). Notably, the 2D BeO sheets are all indirect bandgap insulators. The stacking patterns, orders and the number of stacked layers do not affect the valence band maximum (K points) of band structure in the 2D BeO sheet, which differs from the behavior observed in TMDs[7].

Clearly, there are also significant differences in the SHG of 2D BeO sheets with different layers and stacking patterns. Figure 2(b) indicates that AA(A..) stacking is the optimal stacking pattern to promote SHG in 2D BeO sheets. Notably, SHG can be generated regardless of the parity of the layer number of the 2D BeO sheets, which contrasts with previously observed phenomena where the SHG of even-layered TMDs is negligible[7]. For 2D BeO sheets in the same stacking patterns, BL-AA and TL-AAA stacking have stronger sheet SHG coefficients than the BeO ML owing to a decrease in the band gap (Figure 2(b)). The BeO BL-AA has larger $E_g$ (6.67 eV) and stronger sheet SHG coefficients ($\chi_{22}^{(2)}$(2D) = 14.39 Å·pm/V) compared to those of KBBF-like 2D AlBeBO$_3$F$_2$ ($E_g$ = 5.2 eV; $\chi_{11}^{(2)}$(2D) = 11 Å·pm/V)[48]. The TL-AAA BeO exhibits a remarkable SHG effect ($E_g$ = 6.61 eV; $\chi_{22}^{(2)}$(2D) = 21.40 Å·pm/V) slightly lower than that in h-BN ($E_g$ = 6.00 eV; $\chi_{22}^{(2)}$(2D) = 60.80 Å·pm/V)[4b, 4c] Among 2D BeO sheets in the different stacking patterns, TL-ACE and TL-AAB have the same band gap (6.63 eV), but the SHG coefficients of the TL-ACE ($\chi_{22}^{(2)}$(2D) = 19.26 Å·pm/V) is 2.78 times greater than that of TL-AAB ($\chi_{22}^{(2)}$(2D) = 6.96 Å·pm/V). In the TL structures of 2D BeO sheets, the SHG coefficients of TL-AAA, TL-AAC, TL-AAE, and TL-ACE are significantly larger −about three times greater−than those of other stacking types. It can be observed that all these remaining trilayer structures are consist of non-centrosymmetric ML A and centrosymmetry BL configurations, resulting in SHG coefficients comparable to that in BeO ML. As mentioned earlier, the magnitude of SHG coefficients in 2D BeO sytem is either independent of or sensitive to the number of layers, depending on the stacked patterns.

As shown in Figure 3, the layer-dependent SHG effects in 2D BeO sheets are investigated. As the layer number ($N$) of BeO layers increases, $E_g$ of 2D BeO sheet in AA stacking order slightly decreases from 6.86 eV to 6.67 eV ($N$ = 2), then to 6.61 ($N$ = 3), and ultimately to 6.47 eV ($N$ = ∞), which is consistent with those of other 2D materials[49] (Figure 3(d)). When comparing 2D BeO materials with other NLO bulk materials, one can define an effective thickness $L_z^{eff}$. We employed three models to determine the effective thickness of the 2D BeO sheet: electrostatic potential thickness (EP), van der Waals thickness (vdW)[10] methods, and interlayer distance (ID)[30]. We define the difference between the upper and lower layers of the 2D BeO material where the electrons are absent, as the effective thickness $L_{EP}^{eff}$, as shown in Figure 3(a). This represents the point at which the electrons reach the vacuum energy level. Figure 3(b) illustrates the effective thickness $L_{vdW}^{eff}$, which includes the thickness of 2D material ($d_{2D}$) and the van der Waals thickness on two sides of the material (usually approximated by ~3.4 Å)[10, 18], such that $L_{vdW}^{eff} = 3.4 \times 2 + d_{2D}$. Additionally, Figure 3(a) depicts effective thickness $L_{ID}^{eff} = n \times d_{ID}$, where $n$ is layer number in 2D BeO sheet and $d_{ID}$ is interlayer distance in the bulk BeO materials in AA stacking[30]. The effective thickness of 2D BeO sheet as a function of number of layers ($N$) in AA stacking order is shown in Figure 3(e). The order of effective thickness is $L_{ID}^{eff} < L_{vdW}^{eff} < L_{EP}^{eff}$.

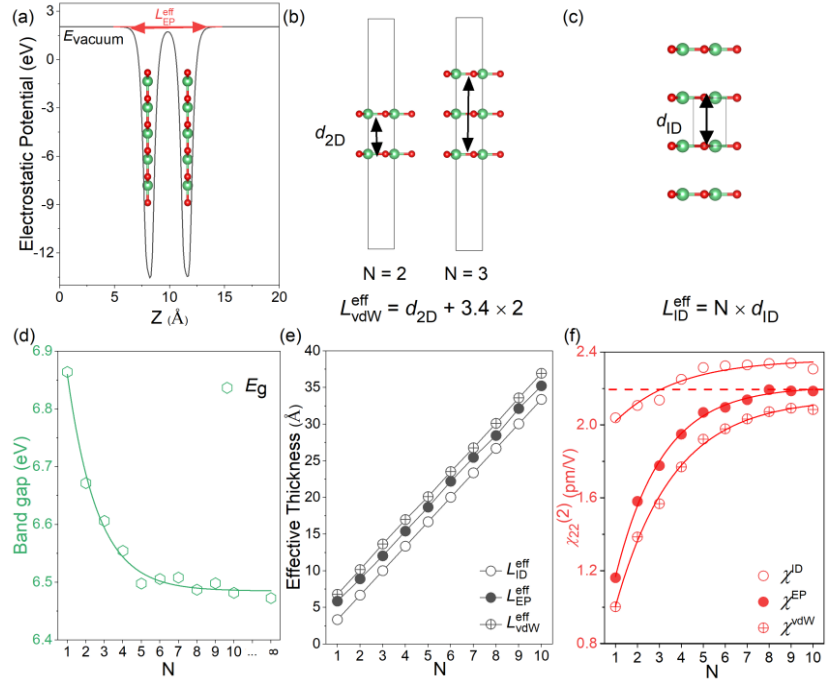

Figure 3 The effective thickness $L_z^{eff}$ of 2D BeO system defined using (a) electrostatic potential (EP), (b) van der Waals thickness (vdW) and (c) interlayer distance (ID) methods. Band gap ($E_g$) (d), effective thickness $L$ and SHG coefficients $\chi_{22}^{(2)}$ as function of number layers ($N$) in AA stacking order. The red dashed line represents the $\chi_{22}^{(2)}$ of the BeO bulk in AA stacking.

Obviously, the existence of inversion asymmetry is independent of $N$ in 2D BeO sheets due to its AA stacking order. Unlike the experimental observation that the SHG coefficients of $MoS_2$[7] oscillate with increasing $N$, the relationship between SHG and $N$ for 2D BeO sheets satisfy an exponential relation because of weak interlayer coupling, as shown in Figure 3(d). The SHG coefficients of 2D BeO sheets gradually increases with the increase $N$ and approaches a stable constant. The SHG coefficients obtained using the effective thickness $L_{EP}^{eff}$ are slightly comparable to that of the BeO bulk phase in the AA stacking, (2.19 pm/V) which is about twice that of the well-known KBBF crystal ($\chi_{11}^{(2)}$ = 0.94 pm/V)[50] (Table S2). This suggests that the effective thickness defined by the EP method is particularly suitable for 2D BeO sheets. Consequently, such a large static SHG susceptibility indicates a strong frequency-doubling effect in 2D BeO sheets, making them promising candidates for ultrathin DUV NLO devices and spectroscopies.

**2.3. Strain dependence of SHG response of 2D BeO sheet.**

Strain inevitably occurs when integrating materials into devices, and it can modulate the electronic and optical properties of the material. Therefore, it is essential to evaluate the strain-dependent band gap and SHG coefficients in 2D BeO sheets. As shown in the Figure 4(a), (b), and (c), the band gap of the 2D BeO sheet exhibits a monotonically decreasing trend with decreasing compressive strain or increasing tensile strain. The effects of stacking patterns on the variation of the band gap are minimal, with changes of only 3.2% for BL-AC, 4% for BL-AA, 4.3% for TL-AAC, 4.5% for TL-AAE, 4.6% for TL-ACE and 4.8% for TL-AAA within stress ($\varepsilon$) range of 5% ≤ $\varepsilon$ ≤ 5% (Figure S7). However, it is noteworthy that the change in band gap increases the numbers of stacking layers with values of 2.5% for ML, 4% for BL-AA, and 4.8% < TL-AAA under biaxial strain in 2D BeO sheets (Figure S6). Undoubtedly, the strain-induced changes in the atomic distances[44], the in-plane stiffness $C$ and Poisson's ratio $v$ (Figure S7) modulate the band gap, as there is no significant charge transfer within the stress range (Figure S8).

It is worth noting that abundant tunability of SHG is achieved in the 2D BeO sheet under biaxial strain. 2D BeO sheets in ML, BL-AA stacking, and TL-AAA stacking exhibit a positive correlation as the biaxial strain varies within ± 5%. Specifically, their SHG reached 7.08, 14.93, 22.77 Å·pm/V at 5% strain, respectively, as illustrated in Figure 4(d). The strain-induced SHG effects in these 2D BeO sheets arise from an increase in polarization due to the asymmetry sublattices of Be and O atoms as the strain amplitude increases. Importantly, the change of SHG response increases with the number of layers (ML: 3.9% < BL-AA: 5.5% < TL-AAA: 7.5%) in Figure S6(a). This can be attributed to the increased thickness of the 2D BeO sheets, which amplifies the strain amplitude and consequently leads to larger SHG effects. Thus, these 2D BeO sheets would offer unprecedented tunability in mechanical-sensitive nano-optical devices and flexible devices with strain-modulated SHG effect.

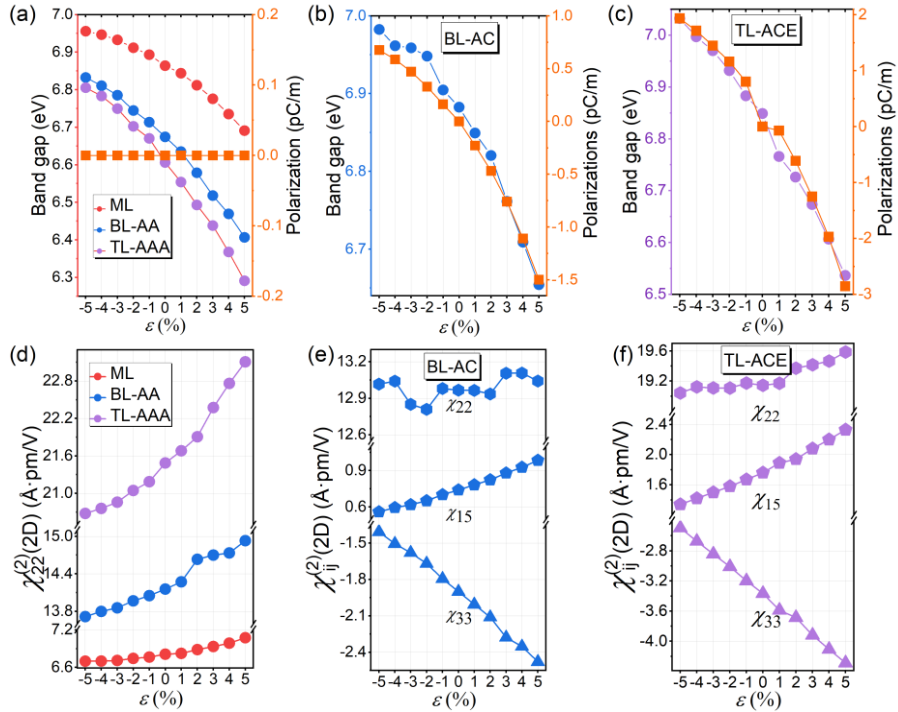

Figure 4 The dependences of band gap, out-of-plane polarizations and SHG effect under biaxial strain in 2D BeO sheets.

More significantly, the diverse and rich modulation effect on the SHG of 2D BeO sheets with *P*3*m*1 symmetry under biaxial strain for stacking types other than AA stacking are illustrated in Figure 4(e), (f) and Figure S9. Because the lack of mirror symmetry puts these 2D BeO sheets in the $C_{3v}$ point group, which exhibit both in-plane and out-of-plane SHG susceptibilities. Taking 2D BL-AC and TL-ACE BeO sheets as examples, intriguingly, except for the in-plane $\chi_{16}^{(2)}$, $\chi_{21}^{(2)}$, and $\chi_{22}^{(2)}$, five additional out-of-plane elements $\chi_{15}^{(2)}$, $\chi_{24}^{(2)}$, $\chi_{31}^{(2)}$, $\chi_{32}^{(2)}$, and $\chi_{33}^{(2)}$ are allowed by the breaking of mirror symmetry. Meanwhile, these elements satisfy the relationship: $\chi_{16}^{(2)} = \chi_{21}^{(2)} = -\chi_{22}^{(2)}$, $\chi_{15}^{(2)} = \chi_{24}^{(2)} = \chi_{31}^{(2)} = \chi_{32}^{(2)}$. Under in-plane biaxial strain (-5% ≤ $\varepsilon$ ≤ 5%), the out-of-plane $\chi_{15}^{(2)}$(2D) and $\chi_{33}^{(2)}$(2D) values in these 2D BL-AC, TL-AAC, TL-AAE, and TL-ACE BeO sheets not only demonstrate negatively/positively correlated tendency for the out-of-plane polarization (Figure 4(b) and (c)), but also show exceptional ~30% change (Figure S10). This is because these 2D BeO sheets exhibit pronounced out-of-plane polarization that varies with strain (unlike the AA-stacked 2D BeO as shown in Figure 4(a)). In other words, strain-mediated interlayer coupling and dipole-dipole interaction contribute to the strong vertical NLO response components. Interestingly, within the same $\varepsilon$ ranges, the in-plane $\chi_{22}^{(2)}$(2D) values for 2D BeO sheets in AC-, AAC-, AAE- and ACE-stacking demonstrate robust SHG effects, changing only slightly by 1.5%, 3.9%, 3.9% and 2.3%, respectively. This behavior is attributed to the strong covalent bonds in plane, which is markedly different from that observed in conventional 2D $MoS_2$[4c, 30]. Thus, such remarkable sensitivity of out-of-plane SHG to compressive or tensile stress and unexpected robustness of SHG against large strains show great potential for applications in mechanical-sensitive nano-optical devices under extreme conditions.

**2.4. Angular dependence of SHG response of the twisted BeO bilayers.**

Twistable interfaces play critical role in tunable crystal symmetry[6] and optical nonlinearities[51] of 2D van der Waals materials and heterostructures. To illustrate the effect of stacking orientation on their SHG, we chose 9 structures for angles within the range of two representative stacking orientations, the nearly parallel (θ = 0°) and antiparallel (θ = 60°) configurations. Before investigating the SHG of the BeO BL with the twist angle ranging from 0° to 60°, we should explore the geometrical structure firstly. The space group of the twist BeO BL with θ = 13°, 22°, 32° and 42° is 150 (*P*321), while the θ = 28°, 38° and 47° has the space group *P*312 (no. 149). They are all belong to non-centrosymmetric $D_3$ point group and thus allows strong

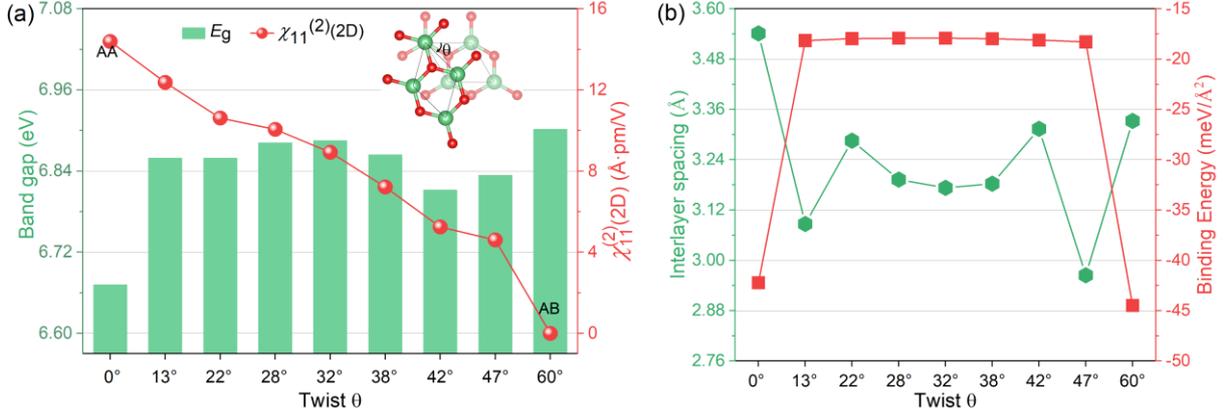

Figure 5. Calculated band gap $E_g$, SHG coefficients $\chi^{(2)}_{11}$(2D) as well as the twist angle. (b) Calculated interlayer spacing between the top and bottom MLs and binding energy in twisted BeO BLs.

SHG. As shown in Figure 5 (a), as the twist angle increases, the SHG gradually decreases, and the bandgap oscillates in the range of 6.53-6.72 eV. The interface angle has a relatively large effect on the band structure (especially for valence band as shown in Figure S11) and a relatively small effect on the band gap, which is slowly changed by ~3.5% ( Figure S12). The twisted BL with θ = 0° (AA stacking configuration) has the strongest sheet SHG susceptibility, which is twice than that of the BeO ML. When the angle is 38°, the SHG susceptibility in the twisted BL is comparable to that of the ML, while the band gap of the twisted BeO BL(6.70 eV) is smaller than that of the BeO ML (6.86 eV) . For the twisted BL with θ = 60° (AB stacking configuration) with inversion symmetry, the SHG signal is completely suppressed. Whereas, the sheet SHG susceptibility of twisted BeO BLs decreases with the increasing of twist angles, revealing a high tunability in the NLO effects[51], which can be modulated as large as ~70% ( Figure S12). This is consistent with the experimentally observed pattern of SHG variation with angle in twisted $WSe_2$/$WSe_2$ BLs[52].

Next, the angle-dependent mechanism of twisted BeO BLs was explored as shown in Figure 5(b). Normally, the interlayer coupling generally exists in the twisted BeO BLs, regardless of the different angles. It is worth to note that interlayer coupling originates from the combination of the decreased interlayer spacing and increased binding energy in the twisted BeO BLs[52]. The strongly suppressed SHG from the twisted BeO BLs with the increasing of twist angles is presumably due to the strong interlayer coupling. As shown in Figure 5, the values of the binding energy/interlayer spacing in twisted BeO BLs are the minimum/maximum for the twist angle at θ = 0° and θ = 60°, and those for other angles are almost the same. In other words, the interlayer coupling is the strongest for these small stacking angles. On one hand, the increase in interlayer coupling strength can effectively increase the band repulsion around the K point (Figure S11), and significantly promote interlayer charge transfer[53] (Figure S13) in twisted BeO BLs for the twist angle in the range of 0° and 60°, which ultimately reduces $\chi_{11}^{(2)}$(2D). On the other hand, electron-hole interactions can also affect the SHG susceptibility[46]. Due to the strong interlayer coupling, the ultrafast photoexcited electrons and holes of the twisted BeO BLs are efficiently transferred from the bottom layer to the top layer (from the top layer to the bottom layer) during the SHG process. They are bound together by fast nonradiative recombination near K point, which greatly reduces the SHG emission of the twisted BeO BLs.

## 3. Conclusions

In summary, our findings provide a theoretical prediction of extraordinary DUV NLO responses in experimentally realizable 2D BeO systems only composed of effective NLO-active [BeO$_3$] unit. An ultrawide band gap (6.86 eV) and a strong SHG responses ($\chi_{22}^{(2)}$(2D) = 6.81 Å·pm/V) simultaneously emerge in 2D BeO ML. Notably, 2D BeO sheets exhibit extremely abundant tunable SHG effect, which can be modulated by the number of stacked layers, stacked patterns, strain and twist angle. The SHG of 2D BeO sheet in AA stacking increases exponentially with the number of stacking layers, converging to that of its bulk phase ($\chi_{22}^{(2)}$ = 2.19 pm/V) when using the effective thickness defined by the electrostatic potential method, which is about twice that of benchmark bulk-KBBF crystal ($\chi_{11}^{(2)}$ = 0.94 pm/V). Those BeO sheets in AC-, AAC-, AAE-, and ACE-stacking exhibit an outstanding 30% variation in in-plane $\chi_{15}^{(2)}$(2D) and $\chi_{33}^{(2)}$(2D) due to out-of-plane polarization and unexpected robustness in-plane $\chi_{22}^{(2)}$(2D) again large strain effects owing to the strong bond in plane under stress stimuli (±5%), differing from other reported 2D NLO materials. Besides, the SHG coefficients of BeO BL decreases with increasing twisted angle due to the strong interlayer coupling, showing extremely high tunable SHG effect. Our theoretical results suggest that 2D BeO sheets are the first synthesized metal oxides as good candidates for application on novel DUV nonlinear integrated nano-optoelectronics. 2D NLO materials constructed only by excellent functional motifs are in anticipation.

## 4. Computational methods

The first-principles calculations are performed based on plane wave pseudopotential method and density functional theory (DFT) using Cambridge Sequential Total Energy Package (CASTEP). The interactions between ionic cores and electrons are described by the norm-conserving pseudopotential. The sheet energy was iterated until a tolerance of $1 \times 10^{-6}$ eV/atom is attained for the electronic relaxation. All the structures are fully relaxed using the Broyden-Fletcher-Goldfarb-Shanno (BFGS) scheme until the residual forces on each atom are less than 0.01 eV Å$^{-1}$. We carefully examine the convergence of properties calculations, especially for the energy cutoff $E_{cut}$, number of empty states $N_{band}$, k-point mesh (k-points), vacuum space $L_z$. It is obvious that $E_{cut}$ = 750 eV, $N_{band}$ = 800, $L_z$ = 30 Å and k-points is 15 × 15 × 1, are enough to converge band gap within 0.02 eV and $\chi^{(2)}_{22}$(2D) within 0.1 Åpm/V as shown in Figure S1 and Figure S2. The semi-empirical Grimme DFT-D correction is adopted to deal with the weak interlayer interactions between the adjacent layers. The energy band structures are calculated by HSE06 hybrid functional calculations based on DFT via the PWmat code[54], a GPU-based code with a plane-wave basis. The plane-wave energy cutoff is set to 50 Ry.

The electric polarizations of the 2D BeO sheets were evaluated by the Berry phase method[55] in the Vienna Ab initio Simulation Package code[56]. A projector augmented wave pseudopotential is used at the general gradient approximation[57] level in the scheme of the Perdew-Burke-Ernzerhof[58] functional. The structure optimization is performed until the energy difference smaller than 10$^{-6}$ eV and the force is less than 10$^{-2}$ eV/Å. The cutoff energy is 520 eV, and the Gamma-centered Monkhorst-Pack[59] k-point is 18 × 18 × 1.

As the thickness of 2D materials is not well defined, the SHG susceptibility $\chi^{(2)}_{ij}$(2D) is obtained by $\chi^{(2)}_{ij}$(2D) = $\chi^{(2)}_{ij}$ (bulk) × $L_z$[60] (unit is Å·pm/V), and the corresponding 3D SHG susceptibility $\chi^{(2)}_{ij}$ is described as $\chi^{(2)}_{ij} = \frac{\chi^{(2)}_{ij}(2D)}{L_z^{eff}}$ (unit is pm/V), where $L_z$ is the thickness in the $c$ direction (the sum of the thickness of the 2D material and the vacuum layer), and $L_z^{eff}$ is the effective thickness, which is determined as three methods as above. To compensate for the underestimated band gap at the PBE level, we use the scissor correction of the band gap difference in the calculations of SHG for higher accuracy.

The interlayer spacing indicates the average perpendicular distance of atoms in adjacent layers, and the binding energy between the two layers is defined by ($E_{TB}$ - $E_T$ - $E_B$)/S, where $E_{TB}$, $E_T$, $E_B$, and S are the total energy of optimized top and bottom layers, the energy of separate top and bottom layers, and the surface area, respectively.

**Supporting Information**

Supporting Information is available from the Wiley Online Library or from the author.

**Acknowledgements**

This work is supported by National Key Research and Development Program of China (2021YFB3601502), the Key Research Program of Frontier Sciences, CAS (ZDBS-LY-SLH035), National Natural Science Foundation of China (22193044, 52403305, 22361132544), CAS Project for Young Scientists in Basic Research (YSBR-024), the Strategic Priority Research Program of the Chinese Academy of Sciences (XDB0880000), Xinjiang Autonomous Region "Tianchi Talent" introduction plan (2024000069), Tianshan Basic Research Talents (2022TSYCJU0001), the Xinjiang Major Science and Technology Project (2021A01001).

Received: ((will be filled in by the editorial staff))

Revised: ((will be filled in by the editorial staff))

Published online: ((will be filled in by the editorial staff))

**References**

[1] a) P. S. Halasyamani, J. M. Rondinelli, *Nat. Commun.* **2018**, 9, 2972; b) G. Shi, Y. Wang, F. Zhang, B. Zhang, Z. Yang, X. Hou, S. Pan, K. R. Poeppelmeier, *J. Am. Chem. Soc.* **2017**, 139, 10645; c) B. B. Zhang, G. Q. Shi, Z. H. Yang, F. F. Zhang, S. L. Pan, *Angew. Chem. Int. Ed.* **2017**, 56, 3916; d) X. F. Wang, Y. Wang, B. B. Zhang, F. F. Zhang, Z. H. Yang, S. l. Pan, *Angew. Chem. Int. Ed.* **2017**, 56, 14119; e) M. Mutailipu, M. Zhang, H. Wu, Z. Yang, Y. Shen, J. Sun, S. Pan, *Nat. Commun.* **2018**, 9, 3089; f) Y. Wang, B. B. Zhang, Z. Yang, S. Pan, *Angew. Chem. Int. Ed.* **2018**, 57, 2150; g) M. Wu, E. Tikhonov, A. Tudi, I. Kruglov, X. Hou, C. Xie, S. Pan, Z. Yang, *Adv. Mater.* **2023**, 35, e2300848.

[2] a) I. Abdelwahab, B. Tilmann, Y. Wu, D. Giovanni, I. Verzhbitskiy, M. Zhu, R. Berté, F. Xuan, L. d. S. Menezes, G. Eda, T. C. Sum, S. Y. Quek, S. A. Maier, K. P. Loh, *Nat. Photon.* **2022**, 16, 644; b) Q. Guo, X. Z. Qi, L. Zhang, M. Gao, S. Hu, W. Zhou, W. Zang, X. Zhao, J. Wang, B. Yan, M. Xu, Y. K. Wu, G. Eda, Z. Xiao, S. A. Yang, H. Gou, Y. P. Feng, G. C. Guo, W. Zhou, X. F. Ren, C. W. Qiu, S. J. Pennycook, A. T. S. Wee, *Nature* **2023**, 613, 53; c) Z. Sun, Y. Yi, T. Song, G. Clark, B. Huang, Y. Shan, S. Wu, D. Huang, C. Gao, Z. Chen, M. McGuire, T. Cao, D. Xiao, W. T. Liu, W. Yao, X. Xu, S. Wu, *Nature* **2019**, 572, 497.

[3] a) X. Yin, Z. Ye, D. A. Chenet, Y. Ye, K. O'Brien, J. C. Hone, X. Zhang, *Science* **2014**, 344, 488; b) Y. Zhou, D. L. Engelberg, *MGE Advances* **2024**, 2, e57.

[4] a) H. Li, J. Min, Z. Yang, Z. Wang, S. Pan, A. R. Oganov, *Angew. Chem. Int. Ed.* **2021**, 60, 10791; b) X. Liu, L. M. Wu, L. Kang, Z. Lin, L. Chen, *Small* **2024**, e2404155; c) L. Kang, X. Liu, Z. Lin, B. Huang, *Phys. Rev. B* **2020**, 102, 205424; d) A. Taghizadeh, K. S. Thygesen, T. G. Pedersen, *ACS Nano* **2021**, 15, 7155.

[5] W. Huang, Y. Xiao, F. Xia, X. Chen, T. Zhai, *Adv. Funct. Mater.* **2024**, 2310726.

[6] F. Yang, W. Song, F. Meng, F. Luo, S. Lou, S. Lin, Z. Gong, J. Cao, E. S. Barnard, E. Chan, L. Yang, J. Yao, *Matter* **2020**, 3, 1361.

[7] Y. Li, Y. Rao, K. F. Mak, Y. You, S. Wang, C. R. Dean, T. F. Heinz, *Nano Lett.* **2013**, 13, 3329.

[8] S. Ahmed, P. K. Cheng, J. Qiao, W. Gao, A. M. Saleque, M. N. Al Subri Ivan, T. Wang, T. I. Alam, S. U. Hani, Z. L. Guo, S. F. Yu, Y. H. Tsang, *ACS Nano* **2022**, 16, 12390.

[9] X. Zhou, J. Cheng, Y. Zhou, T. Cao, H. Hong, Z. Liao, S. Wu, H. Peng, K. Liu, D. Yu, *J. Am. Chem. Soc.* **2015**, 137, 7994.

[10] H. Wang, X. Qian, *Nano Lett.* **2017**, 17, 5027.

[11]  S. J. Varma, J. Kumar, Y. Liu, K. Layne, J. Wu, C. Liang, Y. Nakanishi, A. Aliyan, W. Yang, P. M. Ajayan, J. Thomas, *Adv. Opt. Mater.* **2017**, 5, 1700713.
[12]  K. Younus, Y. Zhou, M. Zhu, D. Xu, X. Guo, A. Ahmed, F. Ouyang, H. Huang, S. Xiao, Z. Chen, J. He, *J. Phys. Chem. Lett.* **2024**, 15, 4992.
[13]  L. Mennel, M. Paur, T. Mueller, *APL Photonics* **2019**, 4, 034404.
[14]  Y. Song, R. Tian, J. Yang, R. Yin, J. Zhao, X. Gan, *Adv. Opt. Mater.* **2018**, 6, 1701334.
[15]  M. M. Petrić, V. Villafañe, P. Herrmann, A. Ben Mhenni, Y. Qin, Y. Sayyad, Y. Shen, S. Tongay, K. Müller, G. Soavi, J. J. Finley, M. Barbone, *Adv. Opt. Mater.* **2023**, 11, 23000958.
[16]  a) Y. Gu, H. Cai, J. Dong, Y. Yu, A. N. Hoffman, C. Liu, A. D. Oyedele, Y. C. Lin, Z. Ge, A. A. Puretzky, G. Duscher, M. F. Chisholm, P. D. Rack, C. M. Rouleau, Z. Gai, X. Meng, F. Ding, D. B. Geohegan, K. Xiao, *Adv. Mater.* **2020**, 32, e1906238; b) J. Yu, X. F. Kuang, J. Z. Li, J. H. Zhong, C. Zeng, L. K. Cao, Z. W. Liu, Z. X. S. Zeng, Z. Y. Luo, T. C. He, A. L. Pan, Y. P. Liu, *Nat. Commun.* **2021**, 12, 1083.
[17]  Y. Shen, Y. Guo, Q. Wang, *Adv. Theory Simul.* **2020**, 3, 2000027.
[18]  C. Hou, Y. Shen, Q. Wang, Y. Kawazoe, P. Jena, *J. Mater. Chem. A* **2023**, 11, 167.
[19]  X. Zou, X. Yuan, L. Liang, F. Tian, Y. Li, Y. Sun, C. Wang, *J. Am. Chem. Soc.* **2024**, 146, 17784.
[20]  J. He, S. H. Lee, F. Naccarato, G. Brunin, R. Zu, Y. Wang, L. Miao, H. Wang, N. Alem, G. Hautier, G.-M. Rignanese, Z. Mao, V. Gopalan, *ACS Photonics* **2022**, 9, 1724.
[21]  C. Y. Zhu, Z. Zhang, J. K. Qin, Z. Wang, C. Wang, P. Miao, Y. Liu, P. Y. Huang, Y. Zhang, K. Xu, L. Zhen, Y. Chai, C. Y. Xu, *Nat. Commun.* **2023**, 14, 2521.
[22]  S. Aoki, Y. Dong, Z. Wang, X. S. W. Huang, Y. M. Itahashi, N. Ogawa, T. Ideue, Y. Iwasa, *Adv. Mater.* **2024**, e2312781.
[23]  K. Bu, T. Fu, Z. Du, X. Feng, D. Wang, Z. Li, S. Guo, Z. Sun, H. Luo, G. Liu, Y. Ding, T. Zhai, Q. Li, X. Lü, *Chem. Mater.* **2022**, 35, 242.
[24]  a) Y. Fang, F. Wang, R. Wang, T. Zhai, F. Huang, *Adv. Mater.* **2021**, 33, 2101505; b) Y. Jia, M. Zhao, G. Gou, X. C. Zeng, J. Li, *Nanoscale Horiz.* **2019**, 4, 1113; c) T. Fu, K. Bu, X. Sun, D. Wang, X. Feng, S. Guo, Z. Sun, Y. Fang, Q. Hu, Y. Ding, T. Zhai, F. Huang, X. Lü, *J. Am. Chem. Soc.* **2023**, 145, 16828; d) J. Fu, N. Yang, Y. Liu, Q. Liu, J. Du, Y. Fang, J. Wang, B. Gao, C. Xu, D. Zhang, A. J. Meixner, G. Gou, F. Huang, L. Zhen, Y. Li, *Adv. Funct. Mater.* **2023**, 34, 2308207.
[25]  T. Su, C. H. Lee, S.-D. Guo, G. Wang, W.-L. Ong, L. Cao, W. Zhao, S. A. Yang, Y. S. Ang, *Mater. Today Phys.* **2023**, 31, 101001.
[26]  X. Li, Y. Guan, X. Li, Y. Fu, *J. Am. Chem. Soc.* **2022**.
[27]  Y. Guo, H. Zhu, Q. Wang, *J. Phys. Chem. C* **2020**, 124, 5506.
[28]  Y. Wang, J. Xiao, T.-F. Chung, Z. Nie, S. Yang, X. Zhang, *Nat. Electron.* **2021**, 4, 725.
[29]  J. Liang, J. Zhang, Z. Li, H. Hong, J. Wang, Z. Zhang, X. Zhou, R. Qiao, J. Xu, P. Gao, Z. Liu, Z. Liu, Z. Sun, S. Meng, K. Liu, D. Yu, *Nano Lett.* **2017**, 17, 7539.
[30]  Q. Wu, F. Liang, L. Kang, J. Wu, Z. Lin, *ACS Appl. Mater. Interfaces* **2022**, 14, 9535.
[31]  G. Wang, X. Marie, I. Gerber, T. Amand, D. Lagarde, L. Bouet, M. Vidal, A. Balocchi, B. Urbaszek, *Phys. Rev. Lett.* **2015**, 114, 097403.
[32]  a) K. Zhou, G. Shang, H. H. Hsu, S. T. Han, V. A. L. Roy, Y. Zhou, *Adv. Mater.* **2023**, 35, e2207774; b) X. Feng, R. Cheng, L. Yin, Y. Wen, J. Jiang, J. He, *Adv. Mater.* **2024**, 36, e2304708.
[33]  A. Tudi, S. Han, Z. Yang, S. Pan, *Inorg. Chem. Front.* **2021**, 8, 4791.
[34]  A. Continenza, R. M. Wentzcovitch, A. J. Freeman, *Phys. Rev. B* **1990**, 41, 3540.

[35] a) H. Zhang, M. Holbrook, F. Cheng, H. Nam, M. Liu, C.-R. Pan, D. West, S. Zhang, M.-Y. Chou, C.-K. Shih, *ACS Nano* **2021**, 15, 2497; b) L. Wang, L. Liu, J. Chen, A. Mohsin, J. H. Yum, T. W. Hudnall, C. W. Bielawski, T. Rajh, X. Bai, S. P. Gao, G. Gu, *Angew. Chem. Int. Ed.* **2020**, 59, 15734.
[36] W. Wu, P. Lu, Z. Zhang, W. Guo, *ACS Appl. Mater. Interfaces* **2011**, 3, 4787.
[37] H. Liu, V. Ksenevich, J. Zhao, J. Gao, *Phys. Chem. Chem. Phys.* **2023**, 25, 8853.
[38] C. Xia, W. Li, D. Ma, L. Zhang, *Nanotechnology* **2020**, 31, 375705.
[39] A. Islam, M. S. Islam, N. Z. Mim, M. S. Akbar, M. S. Hasan, M. R. Islam, C. Stampfl, J. Park, *ACS Omega* **2022**, 7, 4525.
[40] B. Mortazavi, F. Shojaei, T. Rabczuk, X. Zhuang, *Flatchem* **2021**, 28, 100257.
[41] K. B. Joshi, R. Jain, R. K. Pandya, B. L. Ahuja, B. K. Sharma, *J. Chem. Phys.* **1999**, 111, 163.
[42] a) C. Xie, E. Tikhonov, D. Chu, M. Wu, I. Kruglov, S. Pan, Z. Yang, *Sci. China Mater.* **2023**, 66, 4473; b) M. Wu, J. Feng, C. Xie, A. Tudi, D. Chu, J. Lu, S. Pan, Z. Yang, *ACS Appl. Mater. Interfaces* **2022**, 14, 39081; c) Z. Lin, Z. Wang, C. Chen, I. P. Wu, M.-H. Lee, *Chem. Phys. Lett.* **2004**, 399, 125.
[43] a) Z. Li, W. Jin, F. Zhang, Z. Chen, Z. Yang, S. Pan, *Angew. Chem. Int. Ed.* **2022**, 61, e202112844; b) B. Zhang, E. Tikhonov, C. Xie, Z. Yang, S. Pan, *Angew. Chem. Int. Ed.* **2019**, 58, 11726.
[44] Y. Ge, W. Wan, Y. Ren, F. Li, Y. Liu, *Appl. Phys. Lett.* **2020**, 117, 123101.
[45] W. Götze, H. Wagner, *Physica* **1965**, 31, 475.
[46] M. Grüning, C. Attaccalite, *Phys. Rev. B* **2014**, 89, 0811202.
[47] Y. Rouzhahong, B. Zhang, A. Abudurusuli, S. Pan, Z. Yang, *Inorg. Chem.* **2019**, 58, 7715.
[48] G. Yang, K. Wu, *J. Phys. Chem. C* **2018**, 122, 7992.
[49] A. Chaves, J. G. Azadani, H. Alsalman, D. R. da Costa, R. Frisenda, A. J. Chaves, S. H. Song, Y. D. Kim, D. He, J. Zhou, A. Castellanos-Gomez, F. M. Peeters, Z. Liu, C. L. Hinkle, S.-H. Oh, P. D. Ye, S. J. Koester, Y. H. Lee, P. Avouris, X. Wang, T. Low, *npj 2D Mater. and Appl.* **2020**, 4, 41699.
[50] C. T. Chen, G. L. Wang, X. Y. Wang, Z. Y. Xu, *Appl. Phys. B* **2009**, 97, 9.
[51] M. Xu, H. Ji, M. Zhang, L. Zheng, W. Li, L. Luo, M. Chen, Z. Liu, X. Gan, X. Wang, W. Huang, *Adv. Mater.* **2024**, 2313638.
[52] Y. Yuan, P. Liu, H. Wu, H. Chen, W. Zheng, G. Peng, Z. Zhu, M. Zhu, J. Dai, S. Qin, K. S. Novoselov, *ACS Nano* **2023**, 17, 17897.
[53] D. Li, W. Xiong, L. Jiang, Z. Xiao, H. Rabiee Golgir, M. Wang, X. Huang, Y. Zhou, Z. Lin, J. Song, S. Ducharme, L. Jiang, J.-F. Silvain, Y. Lu, *ACS Nano* **2016**, 10, 3766.
[54] a) W. Jia, Z. Cao, L. Wang, J. Fu, X. Chi, W. Gao, L.-W. Wang, *Comput. Phys. Commun.* **2013**, 184, 9; b) W. Jia, J. Fu, Z. Cao, L. Wang, X. Chi, W. Gao, L.-W. Wang, *J Comput Phys* **2013**, 251, 102.
[55] R. D. King-Smith, D. Vanderbilt, *Phys. Rev. B: Condens. Matter.* **1993**, 47, 1651.
[56] G. Kresse, J. Furthmuller, *Phys. Rev., B Condens. Matter.* **1996**, 54, 11169.
[57] J. P. Perdew, K. Burke, M. Ernzerhof, *Phys. Rev. Lett.* **1996**, 77, 3865.
[58] P. E. Blochl, *Phys. Rev. B: Condens. Matter Mater. Phys.* **1994**, 50, 17953.
[59] H. J. Monkhorst, J. D. Pack, *Phys. Rev. B* **1976**, 13, 5188.
[60] A. Strasser, H. Wang, X. Qian, *Nano Lett.* **2022**, 22, 4145.

# Table of Contents Entry

**Two-Dimensional Graphene-like BeO Sheet: A Promising Deep-Ultraviolet Nonlinear Optical Materials System with Strong and Highly Tunable Second Harmonic Generation**

*Linlin Liu, Congwei Xie*, Abudukadi Tudi, Keith Butler and Zhihua Yang**

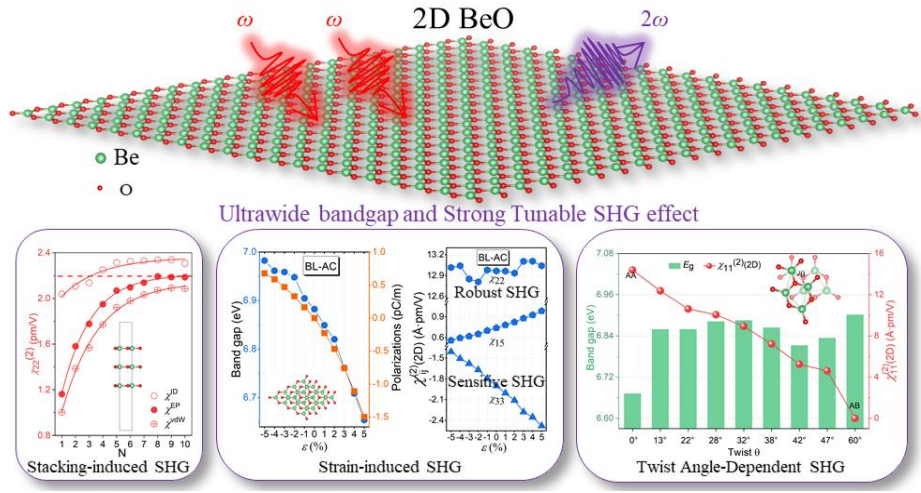

We showed the second harmonics generation in 2D graphene-like BeO with a deep-ultraviolet band gap, explored how the introduction of stacking, strain and twisting degree of freedom in 2D BeO sheets can affect their nonlinear optical behavior.